\documentclass[12pt]{article}

\renewcommand{\title}[1]{\null\vspace{25mm}
\noindent{\Large{\bf #1}}\vspace{10mm}
}
\newcommand{\authors}[1]{\noindent{\large #1}\vspace{20mm}
    }
\newcommand{\address}[1]{{\center{\noindent #1\vspace{10mm}}
    }}
\renewcommand{\abstract}[1]{\vspace{17mm}
\noindent{\small{\em Abstract.} #1}\vspace{2mm}
   }     
 
\newcommand{\mnr}{{\mu\nu\rho}}
\newcommand{\mn}{{\mu\nu}}

\newcommand{\nm}{\nonumber}
\newcommand{\us}{\underline{s}}
\newcommand{\be}{\begin{equation}}
\newcommand{\ee}{\end{equation}}
\newcommand{\ba}{\begin{array}}
\newcommand{\ea}{\end{array}}
\newcommand{\bea}{\begin{eqnarray}}
\newcommand{\eea}{\end{eqnarray}}

\newcounter{saveeqn}
\newcommand{\alpheqn}{
        \setcounter{saveeqn}{\value{equation}}
        \stepcounter{saveeqn}
        \setcounter{equation}{0}
        \renewcommand{\theequation}{\mbox{\arabic{saveeqn}-\alph{equation}}}
        }
\newcommand{\reseteqn}{
        \setcounter{equation}{\value{saveeqn}}
        \renewcommand{\theequation}{\arabic{equation}}
        }

\begin{document}   \setcounter{table}{0}

\begin{titlepage}
\begin{center}
\hspace*{\fill}{{\normalsize \begin{tabular}{l}
                              {\sf hep-th/9909166}\\    
                              {\sf REF. TUW 99-16}\\
                              \end{tabular}   }}

\title{Remarks on Topological SUSY in six-dimensional TQFTs.}

\authors {H.~Ita$^1$, K.~Landsteiner$^2$, T.~Pisar$^3$, J.~Rant$^4$ and M.~Schweda$^5$}    \vspace{-20mm}
       
\address{Institut f\"ur Theoretische Physik,Technische Universit\"at Wien\\
      Wiedner Hauptstra\ss e 8-10, A-1040 Wien, Austria}
\footnotetext[1]{e-mail: ita@hep.itp.tuwien.ac.at}
\footnotetext[2]{Work supported by the "Fonds zur F\"orderung der Wissenschaflicher Forschung", under Project Grant Number P11654-PHY.}
\footnotetext[3]{Work supported by the "Fonds zur F\"orderung der Wissenschaflicher Forschung", under Project Grant Number P11582-PHY.}
\footnotetext[4]{Work supported by the "Fonds zur F\"orderung der Wissenschaflicher Forschung", under Project Grant Number P11354-PHY.}
\footnotetext[5]{email: schweda@hep.itp.tuwien.ac.at}       
\end{center} 
\thispagestyle{empty}
\abstract{We establish the existence of the topological vector supersymmetry
in the six dimensional topological field theory for two-form fields introduced
by Baulieu and West. We investigate the relation of these symmetries to
the twist operation for the $(2,0)$ supersymmetry and comment on their 
resemblance to the analogous symmetries in topological Yang-Mills theory.}
\end{titlepage}

\section {Introduction}

There has been recently some interest in higher dimensional topological 
field theories. In particular Baulieu and West studied a six dimensional
topological model of two-form fields in \cite{Baulieu:1998xd}. They also
discussed the BRST quantization of this model and a new kind of
topological twist operation specific to $(2,0)$ supersymmetry in six
dimensions. The aim of this paper is to elaborate on their work and investigate
closer the BRST-structure of the model and its relation to a twist of
the $(2,0)$ supersymmetry. More specifically we are looking for a completion
of the topological supersymmetry. In \cite{Baulieu:1998xd} the authors
showed that the part of the BRST-operator that encodes the topological
shift symmetry can be thought of being obtained by twisting the $(2,0)$
Poincar{\'e} supersymmetry algebra. The complete BRST operator is built
up by promoting also the usual two-form field gauge symmetries to a 
BRST-symmetry. All this structure is completely parallel to what happens
in four dimensional topological Yang-Mills theory \cite{Witten:1988ze,Baulieu:1988xs}.

In the latter theory it has been shown that there also exists a fermionic
vectorial supersymmetry whose anti-commutator with the BRST-operator closes
onto translation \cite{Brandhuber:1994uf}. 
Of course, the existence of this additional symmetry can to some extent be motivated from the fact that
the twist of the $N=2$ supersymmetry in four dimensions also gives rise
to vectorial supersymmetric partner of the BRST-charge \cite{Fucito:1997xm}.
It seems therefore natural to ask also for such a vector supersymmetry
in the topological model of \cite{Baulieu:1998xd}. It is this question
which we will investigate in this paper. 

The paper is organized as follows. In section two we will present the 
model and discuss the BRST action on the gauge field sector. In section
 three we present the gauge fixing of the shift
 symmetry $Q$ and the symmetry $s$. Additionally, we rewrite the $Q$-fixed action in
 a new basis for later use. In the next
section we establish the existence of the topological supersymmetry and discuss 
the twist operation
of the $(2,0)$ supersymmetry. We obtain the $Q$-operator related
to the topological shift symmetry and also a vector operator.
In this context we point out that the form of the vectorial transformation laws are in 
close analogy to the vectorial transformation laws obtained by twisting
the four dimensional $N=2$ Super Yang Mills theory. Concluding, we state why
the transformations derived from the twisting procedure do not close onto translations modulo 
gauge-transformations and e.o.m.. 

\section {Gauge field and ghosts}

For the construction of a TQFT in six dimensions one needs two-form fields $B_2=\frac{1}{2}B_\mn dx^\mu \wedge dx^\nu$ and ${}^cB_2=\frac{1}{2}{}^cB_\mn dx^\mu \wedge dx^\nu$ constrained by the self-duality conditions between these two independent real two forms
\begin{equation}
        \partial_{[\mu}B_{\nu\rho]}+\frac{1}{6}\varepsilon_{\mu\nu\rho\alpha\beta\gamma}\partial^{[\alpha}{}^c\!B^{\beta\gamma]}=0,
\label{self}
\end{equation}
in order to get a topological invariant action $\Sigma_{cl}$ which does not vanish trivially
\begin{equation}
        \Sigma_{cl} = \int F_3 \wedge {}^c\!F_3.
\label {Scl}
\end{equation} 
The corresponding field strength tensors are defined by $F_3=dB_2$ and ${}^c\!F_3=d{}^c\!B_2$, where $d=dx^\mu\partial_\mu$ denotes the exterior derivative. The action (\ref{Scl}) possesses two gauge symmetries 
\begin{equation}
        \delta_{\lambda_2}B_2=\lambda_2 \mbox{ and } \delta_{\lambda_1}B_2=d\lambda_1,
\label{gauge}
\end{equation}
and similar equations for ${}^c\!B_2$. $\lambda_2$ and $\lambda_1$ are a two-form and a one-form characterizing the infinitesimal gauge symmetry.

The usual BRST-quantization demands that these infinitesimal gauge parameters are transformed into ghost fields $\psi_2^1$ and $V_1^1$. 

Corresponding to equ.(\ref{gauge}) the total BRST operator is therefore defined by
\begin{equation}
        \us=Q+s,
\end{equation}
where $Q$ represents the topological shift symmetry \cite{Witten:1988ze} and $s$ is the usual BRST-symmetry induced by the gauge freedom of the model. Generalizing the concepts of \cite{Bertlmann} the total BRST-symmetry is constructed by the so called horizontality conditions   
\begin{eqnarray}
        \left(\us+d\right) \left( B_2+V^1_1+m^2\right) & =  &F_3+\psi^1_2+\phi^2_1+\phi^3, 
\label{hor1}
\end{eqnarray}
which implies the horizontality condition
\begin{eqnarray}
        \left(\us+d\right) \left(F_3+\psi^1_2+\phi^2_1+\phi^3 \right) & = & 0,
\label {hor2}
\end{eqnarray}
by virtue of the nilpotency of $(\us+d)$. Equ.(\ref{hor1}) yields in terms of forms
\begin{eqnarray}
        F_3 &=& dB_2, \nm \\
        \us B_2&=&\psi_2^1-dV_1^1, \nm \\
        \us V_1^1&=& \phi_1^2-dm^2, \nm \\
        \us m^2&=&\phi^3,
\label{erg1}
\end{eqnarray}
whereas (\ref{hor2}) leads to
\begin{eqnarray}
        dF_3&=&0,\nm \\
        \us F_3&=&-d\psi_2^1,\nm \\
        \us\psi_2^1 &=& -d\phi_1^2,\nm \\
        \us\phi_1^2&=&-d\phi^3, \nm \\
        \us\phi^3&=&0.
\label{erg2}
\end{eqnarray}

The transformations in component notation for $B_\mn, \psi^1_\mn, V_\mu^1,\phi^2_\mu,m^2,\phi^3$ are deduced from (\ref{erg1}) and (\ref{erg2}):
\begin{eqnarray}
        \us B_{\mu\nu} & = & \psi^1_{\mu\nu}+2\partial_{[\mu}V^1_{\nu]}, \nm \\
        \us\psi^1_{\mu\nu} & = & -2\partial_{[\mu}\phi^2_{\nu]},\nm \\ 
        \us V^1_{\mu} & = & \phi^2_{\mu}-\partial_{\mu}m^2, \nm \\
        \us\phi^2_{\mu} & = & \partial_{\mu} \phi^3, \nm \\
        \us m^2 & = & \phi^3, \nm \\
        \us \phi^3 & = & 0.
\end{eqnarray}
One identifies the non-vanishing $Q$-symmetry by
\begin{eqnarray}
        Q B_{\mu\nu} & = & \psi^1_{\mu\nu} , \nm \\
        Q\psi^1_{\mu\nu} & = & -2\partial_{[\mu}\phi^2_{\nu]},\nm \\ 
        Q\phi^2_{\mu} & = & \partial_{\mu} \phi^3,
\label{Q}
\end{eqnarray}
and the non-vanishing BRST-symmetry stemming from the gauge freedom is given through
\begin{eqnarray}
        s B_{\mu\nu} & = & 2\partial_{[\mu}V^1_{\nu]}, \nm \\
        s V^1_{\mu} & = & -\partial_{\mu}m^2.
\label{s}
\end{eqnarray}
The total BRST operator $\us$ and the exterior derivative $d$ acting on the above fields are both nilpotent and fulfill therefore 
\begin{equation}
        \left\{\us,d\right\}=d^2=\us^2=0. 
\end{equation}
The dimensions and the corresponding $\Phi\Pi$-charges of the gauge fields and the ghosts are given in table \ref{tab1}.
\begin {table}[ht]
\begin {center}
\begin{tabular}{|r|r|r|r|r|r|r|} \hline
        & $B_{\mu \nu}$ & $V^1_{\mu}$ & $\psi^1_{\mu \nu}$ & $\phi^2_{\mu}$ & $m^2$ & $\phi^3$ \\ \hline
        dim & 2 & 1 & 2 & 1 & 0 & 0 \\ \hline
        $\Phi\Pi$ & 0 & 1 & 1 & 2 & 2 & 3 \\ \hline
\end{tabular}
\caption{Dimensions and Faddeev-Popov charges}
\label{tab1}
\end{center}
\end {table}

\section {Gauge fixing procedure}

\subsection{The shift symmetry}

In order to implement the self-duality condition (\ref{self}) one introduces a general three-form $H_3=\frac{1}{3!}H_\mnr dx^\mu \wedge dx^\nu \wedge dx^\rho$ with its corresponding anti-ghost $\chi^{-1}_3=\frac{1}{3!}\chi^{-1}_\mnr dx^\mu \wedge dx^\nu \wedge dx^\rho$. These fields are members of a BRST-doublet
\begin{equation}
\begin{array}{rclrcl}
        Q\chi^{-1}_{\mu\nu\rho} &=&  H_{\mu\nu\rho},&  QH_{\mu\nu\rho} &=& 0,
\end{array}
\end{equation}
For the gauge-fixing of the $Q$-symmetry one proceeds similarly by an introduction of the anti-ghost fields $\phi^{-2}_\mu, \phi^{-3},X^1$ with their BRST-doublet partners (see also table \ref{topotab})
\begin{equation}
\begin{array}{lcllcl}
        Q\phi^{-2}_{\mu} & = & \eta^{-1}_{\mu}, & Q\eta^{-1}_{\mu} & = & 0, \nm \\
        Q\phi^{-3} & = & \eta^{-2}, & Q\eta^{-2} & = & 0, \nm \\
        QX^1&=&\eta^2, & Q\eta^2&=&0.
\label{Qgauge}
\end{array}
\end{equation}
\begin{table}[ht]
\begin{center}
\begin{tabular}{ccccccc} 
        && $\psi^1_\mn$ \\
        & $\phi^{-2}_\mu$&&$\phi_\mu^2$ \\
        $\phi^{-3}$&&$X^1$&&$\phi^3$
\end{tabular}
\caption{Topological ghosts and anti-ghosts}
\label{topotab}
\end{center}
\end{table}
In the spirit of \cite{Baulieu:1998xd,PiguetSorella} we are now able to fix the topological shift symmetry with its reducible symmetries by the following gauge conditions \`{a} la BRST\footnote{In equ.(\ref{SQ1}) the notation $cc$ means terms identical to the explicit ones, which are obtained by replacing all fields with their counterparts with index ${}^c$.}
\begin{eqnarray}
        \Sigma_{Q} & = & \int d^6x\;\frac{1}{3!} Q\left\{ \chi^{-1\mu\nu\rho}\left(3\partial_{\mu}B_{\nu\rho}+\frac{1}{6}\varepsilon_{\mu\nu\rho\alpha\beta\gamma}3\partial^{\alpha}{}^c\!B^{\beta\gamma}-\frac{\alpha_1}{2}H_\mnr \right)\right\} \nm \\
        & & +\int d^6x\; Q \left\{ \phi^{-2}_{\mu} \partial_{\nu}\psi^{1\mu\nu}+\phi^{-3} \left( \alpha_2 \eta^2+ \partial^{\mu}\phi^2_{\mu}\right)+X^1 \left( \alpha_3 \eta^{-2}+\partial^{\mu}\phi^{-2}_{\mu} \right) -cc \right\}.
\label{SQ1}
\end{eqnarray}
We also included additional couplings with parameters $\alpha_i$ which are consistent with the symmetries, dimensions and $Q_{\Phi\Pi}$-charges. Later on, we will see that the topological vector susy places some restrictions on these parameters. 

Using the definition of $Q$ one can eliminate the field $H_\mnr$ by using its algebraic equation of motion
\bea
        H_\mnr &=& \frac{1}{\alpha_1} \left( 3\partial_{[\mu}B_{\nu\rho]}+\frac{1}{6}\varepsilon_{\mu\nu\rho\alpha\beta\gamma}3\partial^{\alpha}{}^c\!B^{\beta\gamma} \right)
\label{Heom}
\eea
In this way one gets
\begin{eqnarray}
        \Sigma_{Q} & = & \int d^6x\; \left\{ \frac{1}{2\cdot3!\cdot\alpha_1} \left|3\partial_{[\mu}B_{\nu\rho]}+\frac{1}{6}3\varepsilon_{\mu\nu\rho\alpha\beta\gamma}\partial^{\alpha}{}^c\!B^{\beta\gamma}\right|^2 \right. \nm \\
        & & \left. -\frac{1}{3!}\chi^{-1\mu\nu\rho}\left(3\partial_{\mu}\psi^1_{\nu\rho}+\frac{1}{6}\varepsilon_{\mu\nu\rho\alpha\beta\gamma}3\partial^{\alpha}{}^c\!\psi^{1\beta\gamma}\right) \right\} \nm \\
        & & +\int d^6x\; Q \left\{ \phi^{-2}_{\mu} \partial_{\nu}\psi^{1\mu\nu}+\phi^{-3} \left( \alpha_2 \eta^2+ \partial^{\mu}\phi^2_{\mu}\right) +X^1 \left( \alpha_3 \eta^{-2}+\partial^{\mu}\phi^{-2}_{\mu} \right) -cc \right\}.
\label{SQ2}
\end{eqnarray}
In defining
\bea
        \Sigma^1_{Q}&=& \int d^6x\;\frac{1}{3!} \left|3\partial_{[\mu}B_{\nu\rho]}+\frac{1}{6}\varepsilon_{\mu\nu\rho\alpha\beta\gamma}3\partial^{\alpha}{}^c\!B^{\beta\gamma}\right|^2,
\eea
one gets for $\Sigma^1_Q$ the following alternative expression
\bea
        \Sigma^1_Q&=&\int d^6x\; \frac{1}{3!}\left\{\left|3\partial_\mu B_{\nu\rho}\right|^2- \left|3\partial_\mu{}^cB_{\nu\rho}\right|^2+\frac{2}{6}\varepsilon_{\mu\nu\rho\alpha\beta\gamma}3\partial^\mu B^{\nu\rho}3\partial^\alpha {}^cB^{\beta\gamma}\right\}.
\label{Qteil}
\eea
Equ.(\ref{Qteil}) allows to rewrite equ.(\ref{SQ2}) to
\bea
        \Sigma_{Q} & = & \frac{1}{2\alpha_1}\Sigma_Q^1-\int d^6x\; \frac{1}{3!}\left\{\chi^{-1\mu\nu\rho}\left(3\partial_{\mu}\psi^1_{\nu\rho}+\frac{1}{6}\varepsilon_{\mu\nu\rho\alpha\beta\gamma}3\partial^{\alpha}{}^c\!\psi^{1\beta\gamma}\right) \right\} \nm \\
        & & +\int d^6x\; Q \left\{ \phi^{-2}_{\mu} \partial_{\nu}\psi^{1\mu\nu}+\phi^{-3} \left( \alpha_2 \eta^2+ \partial^{\mu}\phi^2_{\mu}\right)+X^1 \left( \alpha_3 \eta^{-2}+\partial^{\mu}\phi^{-2}_{\mu} \right) -cc \right\}.
\eea
Using $\Sigma_{cl}$ of equ.(\ref{Scl}) one can calculate the complete shift symmetry fixed action 
\bea
        \Sigma_{topo} &=& -\frac{1}{\alpha_1} \Sigma_{cl}+\Sigma_Q \nm \\
        & = & \int d^6x\; \left\{ \frac{1}{2\cdot3!\cdot\alpha_1}\left( \left|3\partial_\mu B_{\nu\rho}\right|^2- \left|3\partial_\mu{}^cB_{\nu\rho}\right|^2\right) \right. \nm \\
        & & \left. -\frac{1}{3!}\chi^{-1\mu\nu\rho}\left(3\partial_{\mu}\psi^1_{\nu\rho}+\frac{1}{6}\varepsilon_{\mu\nu\rho\alpha\beta\gamma}3\partial^{\alpha}{}^c\!\psi^{1\beta\gamma}\right)\right\} \nm \\
        & & +\int d^6x\; \left\{ \eta^{-1\mu}\partial^\nu \psi^1_{\mu\nu} + \eta^2 \partial^\mu \phi^{-2}_\mu +\eta^{-2} \partial^\mu \phi^2_\mu - X^1 \partial^\mu \eta^{-1}_\mu -2 \phi^{-2}_\mu\partial_\nu\partial^{[\mu}\phi^{2\nu]} \right. \nm \\
        & & \left. -\phi^{-3}\partial^\mu \partial_\mu \phi^3+(\alpha_2+\alpha_3)\eta^{-2}\eta^2-cc \right\},
\eea
where we also evaluated the $Q$-variations with the help of (\ref{Q}) and (\ref{Qgauge}). Using the algebraic equations of motion for the fields $\eta^{\pm2}$
\bea
        \eta^{\pm 2} &=& -\frac{1}{\alpha_2+\alpha_3} \partial^\mu\phi^{\pm 2}_\mu,
\eea
and with the choice of $\alpha_1=1$ and  $\alpha_2+\alpha_3=1$ one gets finally
\bea
        \Sigma_{topo} & = & \int d^6x\; \left\{ \frac{1}{2\cdot3!}\left( \left|3\partial_\mu B_{\nu\rho}\right|^2- \left|3\partial_\mu{}^cB_{\nu\rho}\right|^2\right) -\frac{1}{3!} \chi^{-1\mu\nu\rho}\left(3\partial_{\mu}\psi^1_{\nu\rho}+\frac{1}{6}\varepsilon_{\mu\nu\rho\alpha\beta\gamma}3\partial^{\alpha}{}^c\!\psi^{1\beta\gamma}\right)\right\}  \nm \\
        & & +\int d^6x\; \left\{ \eta^{-1\mu}\partial^\nu \psi^1_{\mu\nu} - X^1 \partial^\mu \eta^{-1}_\mu -\phi^2_\mu\partial^2 \phi^{-2\mu}+\phi^{3}\partial^2 \phi^{-3}-cc\right\}. 
\label{Actiontopo}
\eea
We would like to stress that our $Q$-fixed action differs from equ.(2.23) in \cite{Baulieu:1998xd}. It seems that this can be traced back to the following. In \cite{Baulieu:1998xd} the authors assumed the fields $\chi^{-1}_\mnr$, $H_\mnr$ to be self-dual (with $^c\chi_\mnr$ and $^cH_\mnr$ antiself-dual). Since in six dimensions the variation with respect to a self-dual field is antiself-dual one would need to introduce cross terms like ($\chi_\mnr {}^c\!H^\mnr -c.c.$ in order to generate an equation of motion like (\ref{Heom}). The introduction of such cross terms is however equivalent to our method of assuming $\chi^{-1}_\mnr$ and $H_\mnr$ not to fulfill (anti)self-duality conditions.\newline
These mixing terms ($\chi^\mnr \partial_\mu\psi_{\nu\rho}$ and $\chi^\mnr \varepsilon_{\mu\nu\rho\alpha\beta\gamma}\partial^{\alpha}{}^c\!\psi^{1\beta\gamma}$) make it impossible to understand the action as the sum of two free six-form multiplets plus one decoupled part of purely topological packages. \newline
The $Q$-transformations after the insertion of the algebraic equations of motion, equal those found in \cite{Baulieu:1998xd}.
\alpheqn
\bea
        Q B_{\mu\nu} & = & \psi^1_{\mu\nu} , \label{Qbegin}\\
        Q\psi^1_{\mu\nu} & = & -2\partial_{[\mu}\phi^2_{\nu]},\\ 
        Q\phi^2_{\mu} & = & \partial_{\mu} \phi^3,\\
        Q\phi^{-2}_{\mu} & = & \eta_{\mu}^{-1},\\
        Q\eta_\mu^{-1}&=&0, \\
        Q\phi^{-3} & = &-\partial^{\mu}\phi^{-2}_{\mu},\\
        QX^1&=&-\partial^{\mu}\phi^{2}_{\mu},\\
        Q\chi^{-1}_{\mu\nu\rho}&=& \left(3 \partial_{[\mu}B_{\nu\rho]}+\frac{1}{6}\varepsilon_{\mu\nu\rho\alpha\beta\gamma}3\partial^{\alpha}{}^c\!B^{\beta\gamma} \right). \label{Qend}
\eea
\reseteqn 
Except for the equation involving $\chi_\mnr^{-1}$ similar transformations are valid for the $cc$-sector.

For later use, we write the action and the $Q$-transformations in a different basis. In order to write the redefinitions as compact as possible we use the shorthand $\Phi=(F,B,\psi^1,\phi^{\pm2},\phi^{\pm3})$:

\begin{eqnarray}
  \label{eq:redefined fields}
\Phi_{\pm}&=&\Phi\pm^{c}\!\Phi,\\
(\chi^{1})^{\pm}&=&\frac{1}{2}(1\pm*)\chi^{-1},
\end{eqnarray}
We denote the (anti-)selfdual parts of a three form by the superscripts $+(-)$, whereas the subscript $+(-)$ , is related to the plus (minus) sign in the field redefinitions in (\ref{eq:redefined fields}).
Rewriting the action (\ref{Actiontopo}) yields:
\begin{eqnarray}
  \label{eq:redefinedaction}
  \Sigma_{topo} & = & \int \; \left\{ -\frac{1}{2\cdot 3!} 3\partial_{[\mu}(B_{+})_{\nu\rho]}3\partial^{\mu}(B_{-})^{\nu\rho}\right.\nm \\
        & & \quad\quad\left.-\frac{1}{3!\cdot 3!}\varepsilon^{\mu\nu\rho\sigma\tau\lambda}(\chi^{-})^{-1}_{\mu\nu\rho} 3\partial_{\sigma}(\psi_{+})_{\tau\lambda}+\frac{1}{3!\cdot 3!}\varepsilon^{\mu\nu\rho\sigma\tau\lambda}(\chi^{+})^{-1}_{\mu\nu\rho} 3\partial_{\sigma}(\psi_{-})_{\tau\lambda}\right\} \nm \\
        & & +\int d^6x\frac{1}{2}\; \left\{ (\eta_{+})^{-1\mu}\partial^\nu(\psi^1_{-})_{\mu\nu} - (X^1_{+}) \partial^\mu (\eta^{-1}_{-})_\mu -(\phi^2_{+})_\mu\partial^2 (\phi^{-2}_{-})^{\mu}+\phi^{3}_{+}\partial^2 \phi^{-3}_{-}\right\} \nm \\
        & & +\int d^6x\frac{1}{2}\; \left\{ (\eta_{-})^{-1\mu}\partial^\nu(\psi^1_{+})_{\mu\nu} - (X^1_{-}) \partial^\mu (\eta^{-1}_{+})_\mu -(\phi^2_{-})_\mu\partial^2 (\phi^{-2}_{+})^{\mu}+\phi^{3}_{-}\partial^2 \phi^{-3}_{+}\right\}. 
\end{eqnarray}
If one collects the fields into two the $+(-)$ sectors as $\{\Phi_{+},(\chi^{-1})^{+}\}$ and $\{\Phi_{-},(\chi^{-1})^{-}\}$ the action can be read as the product of these sectors. Furthermore the $Q$-transformation respects this splitting, i.e. $Q$ does not mix the two sectors:
\begin{eqnarray}
  \label{eq:redefinedQ}
        Q (B_{\pm})_{\mu\nu} & = & (\psi^1_{\pm})_{\mu\nu} ,\\
        Q(\psi^1_{\pm})_{\mu\nu} & = & -2\partial_{[\mu}(\phi^2_{\pm})_{\nu]},\\ 
        Q(\phi^2_{\pm})_{\mu} & = & \partial_{\mu} \phi^3_{\pm},\\
        Q(\phi^{-2}_{\pm})_{\mu} & = & (\eta^{-1}_{\pm})_{\mu},\\
        Q(\eta_{\pm})_\mu^{-1}&=&0, \\
        Q\phi^{-3}_{\pm} & = &-\partial^{\mu}(\phi^{-2}_{\pm})_{\mu},\\
        QX^1_{\pm}&=&-\partial^{\mu}(\phi^{2}_{\pm})_{\mu},\\
        Q\chi^{-1,\pm}_{\mu\nu\rho}&=&(F_{\pm}^{(\pm)})_{\mu\nu\rho}.   
\end{eqnarray}
\subsection{The $s$-symmetry}

\begin{table}[h]
\begin{center}
\begin{tabular}{ccccc} 
        && $B_\mn$ \\
        & $V^{-1}_\mu$&&$V_\mu^1$ \\
        $m^{-2}$&&$n$&&$m^2$
\end{tabular}
\caption{Ghosts and anti-ghosts corresponding to the $s$-symmetry}
\label{gaugetab}
\end{center}
\end{table}In order to be complete we review also the gauge fixing procedure for the $s$-transformation of equ.(\ref{s}) implying the existence of the corresponding anti-ghost fields and multiplier fields (see also table \ref{gaugetab})
\be
\ba{lcllcl}
        \us V^{-1}_{\mu} & = & b_{\mu}, & \us b_{\mu} & = & 0, \\
        \us m^{-2} & = & \beta^{-1}, & \us \beta^{-1} & = & 0,\\
\ea
\ee
where the vector field $V^{-1}_\mu$ exhibits the usual reducible gauge symmetry. This entails the existence of a anti-ghost field $n$ with a multiplier $\beta^1$
\be
\ba{rclrcl}
        \us n&=&\beta^1,& \us\beta^1&=&0.
\ea
\ee
The s-gauge-fixed action becomes therefore
\be
        \Sigma_{gf} =  \int d^6x\; \us \left\{ V^{-1}_\mu \frac{1}{3!}\left(3\partial_\nu B^{\mu\nu}-\frac{\gamma_1}{2}b_\mu\right) + m^{-2}\left(\gamma_2\beta^1+\partial^{\mu} V^1_{\mu}\right)+n\left(\gamma_3\beta^{-1}+\partial^{\mu}V^{-1}_{\mu}\right)-cc \right\}. 
\ee
Choosing $\gamma_1=-1$ and $\gamma_2=\gamma_3=0$ this is the gauge-fixed action of \cite{Baulieu:1998xd}. On the other hand if we integrate out the multiplier fields $b_\mu$ and $\beta^{\pm 1}$ with their algebraic equations of motion
\bea
        b_\mu &=& \frac{1}{\gamma_1}\left(3\partial^\nu B_\mn-\partial_\mu n \right), \nm \\
        \beta^{\pm 1}&=& \mp \frac {1}{\gamma_2-\gamma_3}\partial^\mu V_\mu^{\pm 1},
\eea 
and the selection of $\gamma_1=1$ and $\gamma_2-\gamma_3=-1$ we find
\bea
        \Sigma_{gf} &=&  \int d^6x\; \left\{ \frac{1}{2\cdot3!}\left|3\partial^\nu B_\mn \right|^2-m^{-2}\partial^2 m^2+V_\mu^{-1} \partial^2 V^{1\mu} \right.\nm\\
        && \left.-\frac{1}{2}n \partial^2 n -V^{-1}_\mu\partial_\nu\psi^{1\mu\nu}+m^{-2}\partial^\mu\phi^2_\mu -cc \right\}. 
\label{SBRST}
\eea
By construction, the complete and fully gauge-fixed action 
\begin{equation}
        \Sigma_{total}=-\Sigma_{cl}+\Sigma_{Q}+\Sigma_{gf},
\label{Sgf}
\end{equation} 
is invariant under the action of $\us=s+Q$.

The dimensions and $\Phi\Pi$-charges of the anti-ghost and multiplier fields are given in table \ref{tab2}. Finally we want to mention, that the number of bosonic degrees of freedom equals the number of fermionic degrees of freedom. This terminates the review of the topological model presented in \cite{Baulieu:1998xd}. 

\begin{table}[ht]
\begin{center}
\begin{tabular}{|r|r|r|r|r|r|r|r|r|} \hline
        & $\phi^{-3}$ & $\eta^{-2}$ & $m^{-2}$ & $\beta^{-1}$ & $\phi^{-2}_\mu$ & $\eta^{-1}_\mu$ & $\chi^{-1}_{\mu\nu\rho}$ & $H_{\mu\nu\rho}$ \\ \hline 
        dim &4&4&4&4&3&3&3&3\\ \hline 
        $\Phi\Pi$ &-3&-2&-2&-1&-2&-1&-1&0\\ \hline \hline 
        & $V^{-1}_\mu$ & $b_\mu$ & $L_\mu$ & $\eta^1_\mu$ & $n$ & $\beta^1$&$X^1$&$\eta^2$ \\ \hline 
        dim &3&3&3&3&2&2&2&2\\ \hline 
        $\Phi\Pi$ &-1&0&0&1&0&1&1&2\\ \hline 
\end{tabular}
\caption{Faddeev-Popov charges for the Lagrange multipliers and anti-ghosts.}
\label{tab2}
\end{center}
\end{table}

\section {The topological linear vector supersymmetry}

It is well known that topological field models of Schwarz type are also characterized by the fact that these models posses an additional symmetry: the topological linear vector supersymmetry, which is responsible for the perturbative finiteness of the model. In \cite{Brandhuber:1994uf} is shown that this topological linear vector susy is also present in the topological Yang-Mills theory in four dimensional space-time if one uses a linear gauge condition for the shift ghost $\psi^1_\mu$ instead of the non-linear gauge condition of this Witten-type model. However, one knows that this model is not perturbative finite. The existence of this topological linear vector susy improves considerable the discussion of the algebraic renormalization procedure - leading to an anomaly free theory \cite{Brandhuber:1994uf}. Additionally, one has to stress that in Schwarz type models quantized in the axial gauge the constraints coming from the vector susy allow only tree-graphs which are alway!
 s perturbatively finite.

It seems to be quite natural to investigate the possibility of the existence of this vector susy also for the model under discussion.Usually, the vector susy can be constructed from the fact that $\Sigma_{cl}$ is metric-independent, whereas the metric dependence emerges by the gauge-fixing procedure entailing that the energy-momentum tensor can be written as BRST exact expression \cite{PiguetSorella}
\be
        T_\mn=\us \Lambda_\mn.
\label{T}
\ee
From equ.(\ref{T}) follows in a standard manner in calculating the divergence of the energy-momentum tensor - and in considering\footnote{$\phi_i$ stands collectively for all fields characterizing the topological field model.}
\be
        \int d^6x\;\partial^\mu T_\mn=\int d^6x\;\sum_i\delta_\nu\phi_i\frac{\delta\Sigma}{\delta \phi_i},
\label{ward}
\ee
the existence of the vector susy. Due to the fact that the model in question is a free Abelian one, no external unquantized sources are needed in order to describe the non-linear pieces of $\us$. This implies that the $\delta_\mu$-symmetry is not broken linearly by the quantum fields as it is the case for interacting topological field models. Usually, the BRST-operator and the $\delta_\mu$-operation closes on on-shell on space-time translations 
\be
        \{\us,\delta_\alpha\}=\partial_\alpha + eom.
\label{algebra}
\ee
However, for the model under consideration the Wess-Zumino type algebra (\ref{algebra}) is an off-shell relation. (\ref{algebra}) implies that the $\delta_\mu$-symmetry is in some sense the inverse of the BRST-symmetry. 

For the explicit construction of the vector susy we have chosen a general ansatz which coefficients are determined by the algebra (\ref{algebra}) and the Ward identity (\ref{ward}). This places also restrictions on the gauge parameters.
In particular, we have to set $\alpha_1=\alpha_2=\alpha_3=0$ and $\gamma_i=0$ to obtain a model which is invariant under the vector susy. The transformations of the gauge field and the ghosts then are
\begin{equation}
\begin{array}{lcllcl}
        \delta_{\alpha}B_{\mu\nu} & = & 0, & \delta_{\alpha}V_{\mu}^1 & = & 0 ,\\
        \delta_{\alpha}\psi^1_{\mu\nu} & = & \partial_{\alpha}B_{\mu\nu}, & \delta_{\alpha}\phi^2_{\mu} & = & \partial_{\alpha}V^1_{\mu}, \\
        \delta_{\alpha}\phi^3 & = & \partial_{\alpha}m^2, & \delta_{\alpha}m^2 & = & 0, 
\label{vectorsusy1}
\end{array}
\end{equation}
whereas the anti-ghosts and multipliers transform as
\begin{equation}
\begin{array}{lcllcl}
        \delta_{\alpha}X^1 & = & \partial_{\alpha}n, & \delta_{\alpha}\eta^2 & = & \partial_{\alpha}X^1-\partial_{\alpha}\beta^1, \\
        \delta_{\alpha}V^{-1}_{\mu} & = & \partial_{\alpha}\phi^{-2}_{\mu}, & \delta_{\alpha}b_{\mu} & = & \partial_{\alpha} V^{-1}_{\mu}-\partial_{\alpha}\eta^{-1}_{\mu}, \\
        \delta_{\alpha}n & = & 0, & \delta_{\alpha}\beta^1 & = & \partial_{\alpha}n,\\
        \delta_{\alpha}\phi^{-2}_{\mu} & = & 0, & \delta_{\alpha}\eta^{-1}_{\mu} & = & \partial_{\alpha}\phi^{-2}_{\mu} , \\
        \delta_{\alpha}\chi^{-1}_{\mu\nu\rho} & = & 0, & \delta_{\alpha}H_{\mu\nu\rho} & = & \partial_{\alpha}\chi^{-1}_{\mu\nu\rho},\\
        \delta_{\alpha}m^{-2} & = & -\partial_\alpha \phi^{-3}, & \delta_{\alpha}\beta^{-1} & = & \partial_{\alpha}m^{-2}+\partial_{\alpha}\eta^{-2},\\
        \delta_{\alpha}\phi^{-3} & = & 0, & \delta_{\alpha}\eta^{-2} & = & \partial_{\alpha}\phi^{-3}.
\label{vectorsusy2}
\end{array}
\end{equation}
One deduces that for each BRST-doublet there exists also a corresponding $\delta_\alpha$-doublet. It is straightforward to prove that $\delta_\alpha \left(\Sigma_{cl}+\Sigma_Q+\Sigma_{gf}\right)=0$ is indeed valid. 

\section{Twist}

In a next step we try to reconstruct the topological shift and vector-super\-symmetry transformations of the fields through a twist of a supersymmetry multiplet. Hence, we need a tensor multiplet containing a second-rank antisymmetric tensor gauge field. In six dimensions this is given by a (2,0) tensor multiplet which consists of the desired tensor $\tilde B_\mn$, a real field $\Phi_{ij}$ and a spinor $\kappa_{\alpha i}$.

The spinors in six dimensions transform under $Spin(1,5) \times Spin(5)$, where $Spin(5)$ is isomorphic to $USp(4)$. $USp(4)$ also provides the existence of symplectic Majorana-Weyl spinors. The symplectic Majorana-Weyl condition is 
\begin{equation}
        \kappa^*_{\alpha i}=B_{\dot\alpha}{}^\beta\Omega^{ij}\kappa_{\beta j},
\end{equation}
where $\alpha=1,2,3,4$ are the spinor indices which transform under Weyl-projected $Spin(1,5)$, whereas  $i=1,2,3,4$ denote internal indices transforming under $USp(4)$. The metric $\Omega^{ij}$ is real and antisymmetric and allows raising and lowering of the internal indices by $\kappa^i=\Omega^{ij}\kappa_j$ respectively $\kappa_i=\kappa^j\Omega_{ji}$. The matrix $B$ is unitary and acts like $\psi_{\dot\alpha}=B_{\dot\alpha}^\beta\psi_\beta$ and $\psi^\alpha=\psi^{\dot\beta}B_{\dot\beta}^\alpha$. 

With the constraint $\Omega^{ij}\Phi_{ij}=0$ this yields three respectively five bosonic on-shell degrees of freedom for $\tilde B_\mn$ and $\Phi_{ij}$, and eight fermionic on-shell degrees of freedom for $\kappa_{\alpha i}$ \cite{Bergshoeff:1999my1}.

In order to perform the topological twist \cite{Witten:1994ev} we have to break down $SO(1,5)$ to $SO(5)$, since $SO(5)$ is locally isomorphic to $USp(4)$. Now the twist is done by choosing the diagonal subgroup of the symmetry group $SO(5) \times USp(4)$. The invariant axis under $SO(5)$ will be denoted by $n^{\mu}=(1,\vec{0})$ in the vector representation. For the spin-representation the vector $n^{\mu}$ is translated to $n_{\mu}\left(\bar\gamma^{\mu}\right)^{\alpha\beta}$. The $Spin(5)$ sub-group consists of the rotations of $Spin(1,5)$ that do not change $n_{\mu}\left(\bar\gamma^{\mu}\right)^{\alpha\beta}$. Now technically diagonalizing $SO(5)\times USp(4)$ is done by identifying the respective indices $(\alpha\rightarrow i,...)$. Some care is in order, as a metric, similar to $\Omega_{ij}$ of $USp(4)$ is missing so far.  As a metric of $Spin(5)$  $n_{\mu}\left(\bar\gamma^{\mu}\right)^{\alpha\beta}$ has to be chosen, as it is invariant by definition and any different met!
 ric would further constrain the set of rotations. Identifying indices forces one to identify the matrices as well, so one has to set
\begin{equation}
        n_{\mu}\left(\bar\gamma^{\mu}\right)^{\alpha\beta}\rightarrow n_{\mu} \left(\bar\gamma^{\mu} \right)^{ij}=\Omega^{ij}.
\end{equation}
The supersymmetry action on the components of the complexified multiplet is given by\footnote{The analytic continuation is understood formally since it breaks the Schouten identities which provide the closure of the algebra \cite{Bergshoeff:1999my1}.}

\begin{equation}\begin{array}{rcl}
        \delta \Phi_{jk} & = & \varepsilon^{\alpha}_{[j}\kappa_{\alpha k]}+\frac{1}{4}\Omega_{jk}\Omega^{il}\varepsilon^{\alpha}_{l}\kappa_{\alpha i},\\
        \delta \bar{\Phi}^{jk} & = & -\bar\varepsilon^{\dot\alpha [j}\bar\kappa_{\dot\alpha}^{k]}-\frac{1}{4}\Omega^{jk} \Omega_{il} \bar\varepsilon^{\dot\alpha l}  \bar\kappa^{i}_{\dot\alpha},\\
        \delta \kappa_{\alpha i} & = & {\tilde F}^+_\mnr \left(\gamma^\mnr\right)_{\alpha\dot\beta}\varepsilon^{\dot\beta}_{ i}+\left( \gamma^\mu\partial_\mu\right)_{\alpha\dot\beta}\Phi_{ij}\bar\varepsilon^{\dot\beta j},  \\
        \delta \bar\kappa_{\dot\alpha}^{i} & = & \bar\varepsilon^{\beta i }\bar {\tilde F}^+_\mnr \left(\gamma^\mnr\right)_{\beta\dot\alpha}+\varepsilon^{\beta}_{j}\left( \gamma^\mu\partial_\mu\right)_{\beta\dot\alpha} \bar\Phi^{ij},\\
        \delta \tilde B_{\mu\nu} & = & \bar\kappa_{\dot\alpha}^{i} \left(\gamma_{\mu\nu} \right)^{\;\dot\alpha}_{\dot\beta}\varepsilon^{\dot\beta}_{i}.\\
        \delta (\bar{\tilde{B}})_{\mu\nu} & = & \bar\varepsilon^{\alpha i} \left(\gamma_{\mu\nu} \right)_{\alpha}^{\;\beta}\kappa_{\beta i},
\end{array}\end{equation} 
By an expansion with respect to the gamma matrices the spinors $\kappa_{\alpha i}$ and $\bar\kappa_{\dot\alpha}^i$ can be written as
\begin{equation}\begin{array}{rcl}\label{decomp}
        \kappa_{\alpha i} & = & \left( \gamma^\mnr \right)_{\alpha i} (\chi^{-1,+})_\mnr+\left(\gamma^\mu\right)_ {\alpha i}(\eta_{+}^{-1})_\mu, \\
        \bar\kappa_{\dot\alpha}^{\;i} & = & \left(\gamma^\mn\right)_{\dot\alpha}^{\;i}(\psi^1_{+})_\mn+B_{\dot\alpha}^{\;i} X^{1}_{+},
\end{array}\end{equation}
whereas the scalars are
\begin{equation}
\begin{array}{rr}
        \Phi_{ij} = \left(\gamma^\mu\right)_{ij}(\phi^{-2}_{+})_\mu, & \bar\Phi^{ij}=\left(\bar\gamma^\mu\right)^{ij}(\phi_{+}^2)_\mu.
\end{array}
\end{equation}
The fields $\tilde B_\mn$ and $\bar{\tilde B}_\mn$ recombine to the gauge field $(B_{+})_\mn$
\begin{equation}
        (B_{+})_\mn = -\frac{1}{8}\left(\tilde B_\mn +\bar{\tilde B}_\mn \right).
\end{equation}  
The twist yields for the topological shift symmetry $\tilde Q$
\begin{equation}\begin{array}{rcl}\label{twistq}
        \tilde Q(\phi^{-2}_{+})_\mu & = & -P_\mn\eta^{-1\nu}_{+}, \\
        \tilde Q(\phi^2_{+})_\mu & = & 0,\\
        \tilde Q(\chi^{-1,+})_\mnr & = & -(F_{+})^+_\mnr, \\
        \tilde Q(\eta_{+}^{-1})_\mu & = & 0,\\
        \tilde Q X^1_{+} & = & -\partial_{\mu}(\phi^{2}_{+})^{\;\mu},\\
        \tilde Q(\psi^{1}_{+})_\mn & = & -\partial_{\left[\right.\mu}(\phi^2_{+})_{\nu\left.\right]}, \\ 
        \tilde Q (B_{+})_\mn & = & (\psi_{+}^1)_{\mn}, 
\end{array}\end{equation}
where $P_\mn$ projects to the five dimensional subspace $P_\mn=g_\mn-n_\mu n_\nu$.
It has been emphasized in \cite{Baulieu:1998xd} that one has to add
a ``topological package'' of scalar fields of opposite statistics 
$(\phi^{\pm 3},(\phi^{\pm 2})_{0})$ to match the six dimensionally
covariant form of (\ref{eq:redefinedQ}). We also emphasize that in order
to get the field content of the Q-fixed action in section (3.1) we have
to take also a twisted $(0,2)$ multiplet, which gives us anti-self dual
part.

We also can find a vector transformation from the
twist, which we denote by $\tilde{\delta}_\alpha$\footnote{$\tilde{Q}$ and
$\tilde{\delta}_\alpha$ are obtained by using analogous decompositions as
in (\ref{decomp}) for the supersymmetry parameters. This implies that there
are also some tensorial transformations. We do not further investigate these.}.
%
\begin{equation}\begin{array}{rcl}\label{vectwist}
        \tilde\delta_\alpha(\phi_{+}^{-2})_\mu & = & 0, \\
        \tilde\delta_\alpha (\phi^2_{+})_\mu & = & -2 g^{\nu\rho}P_\mn (\psi^1_{+})_{\alpha\rho}-P_{\alpha\mu} X^1_{+}, \\
        \tilde\delta_\alpha (\psi_{+}^{1})_{\mu\nu} & = &-3!(\bar {F_{+}})^+_{\alpha\mu\nu}, \\
        \tilde\delta_\alpha X^1_{+} & = & 0, \\
        \tilde\delta_\alpha (\eta_{+}^{-1})_\mu & = & - \left( g_{\alpha\mu} \partial^{\rho}(\phi^{-2}_{+})_{\rho}-\partial_{\mu}(\phi^{-2}_{+})_{\alpha}-\partial_{\alpha}(\phi_{+})^{-2}_{\mu}\right), \\
        \tilde\delta_\alpha \chi_\mnr^{-1,+} & = & -\frac{1}{2}\left( g_{\alpha[\mu} \partial_\nu (\phi^{-2}_{+})_{\rho]}-\frac{1}{3!}\varepsilon_{\alpha\mu\nu\rho\kappa\lambda} \partial^{\kappa}(\phi^{-2}_{+})^{\lambda} \right)
, \\
        \tilde\delta_\alpha (B_{+})_\mn & = & 3!(\chi^{-1,+})_{\alpha\mu\nu}-g_{\alpha[\mu} (\eta^{-1}_{+})_{\nu]}\,.
\end{array}\end{equation}

In \cite{Fucito:1997xm} similar results for the shift symmetry and the vector supersymmetry were obtained in the case of Witten's topological Yang-Mills theory by twisting the $N=2$ supersymmetric algebra. However, an analysis of the results of \cite{Fucito:1997xm} shows that they differ from the results given in \cite{Brandhuber:1994uf}. This is due to the fact that the twist procedure starts with a supersymmetric multiplet in the Wess-Zumino-gauge. The super-Poincar\'{e} algebra closes on translations only modulo gauge-transformations and equations of motion. Hence, the twist leads to an algebra that looks like
\bea
        \{Q,\tilde\delta_\alpha\}=\partial_\alpha+\mbox{equations of motion}+\mbox{gauge-transformations},
\label{algebraWZ}
\eea
where we denoted the twisted vector supersymmetry by $\tilde\delta_\alpha$. The model which one obtains, is not totally gauge-fixed, since the gauge-degree of freedom for $A_\mu$ is left \cite{Witten:1988ze}. We expect that the twist in a manifestly $N=2$ supersymmetric formulation would give a result as in
\cite{Brandhuber:1994uf}. In this way the existence of a vector supersymmetry
in the gauge fixed $N=2$ SYM can be motivated. 

Let us contrast this now to our findings in the six dimensional model at hand.
The vector operator (\ref{vectwist}) can be constructed in rather analogous
way as in the four dimensional SYM. Its properties are however rather
different. We find that $\{\tilde{Q},\tilde{\delta}_\alpha\}$ does not close
on translations modulo gauge transformations and equations of motion.
This is because the six dimensional supersymmetry makes use of Schoutens
identities (see e.g. \cite{Bergshoeff:1999my1}) valid for symplectic
Majorana-Weyl spinors. The latter property is however broken by the twist,
as already noted in \cite{Baulieu:1998xd}. In order to connect the twisted
$(2,0)$ multiplet with the field content of the topological model one
has also to introduce the ``topological package'' $(\phi^{\pm 3},(\phi^{\pm 2})_{0})$. One might hope that these fields can compensate the terms breaking the Schoutens identities. However one gets disappointed. It is not possible to incorporate $\phi^{\pm 3}$ such that the algebra closes even on these fields. Lorentz symmetry and ghost number conservation make it impossible.

Therefore, contrary to the case in four dimensions, 
the existence of a topological vector susy we found in the previous chapter seems rather unrelated with 
the fact that part of the BRST-operator can be constructed by a twist of
the six dimensional $(2,0)$ multiplet. 


\providecommand{\href}[2]{#2}\begingroup\raggedright\endgroup

\end{document}